\documentclass[namedreferences]{SolarPhysics}
\usepackage[optionalrh]{spr-sola-addons} 
\usepackage{graphicx}                    
\usepackage{color}                       
\usepackage{url}                         

\begin{document}

\begin{article}

\begin{opening}

\title{Multi-height Scattering Phase Shifts From Sunspots And Phase Differences Between Photospheric Heights With SDO/HMI and AIA Data}

\author{S\'ebastien Couvidat$^{1}$}

\institute{$^{1}$ W.W. Hansen Experimental Physics Laboratory, Stanford University, CA 94305, USA\\}

\date{Received: 5 June 2013}

\runningauthor{S. Couvidat}
\runningtitle{Scattering Phase Shifts From Sunspots And Phase Differences Between Photospheric Heights}

\begin{abstract}
Following Couvidat (2013), we analyze data from the {\it Helioseismic and Magnetic Imager} (HMI) and the {\it Atmospheric Imaging Assembly} (AIA) instruments onboard the \textit{Solar Dynamics Observatory}: Doppler velocity and continuum intensity at 6173 \AA\ as well as intensity maps at 1600 \AA\ and 1700 \AA. Datasets of 14 active regions and four quiet-Sun regions are studied at four heights in the solar photosphere.  
An Hankel-Fourier analysis is performed around these regions of interest.
Outgoing-ingoing scattering phase shifts at a given atmospheric height are computed, as well as ingoing-ingoing and outgoing-outgoing phase differences between two atmospheric heights. The outgoing-ingoing phase shifts produced by sunspots show little dependence on measurement height, unlike the acoustic power reduction measured in Couvidat (2013). Phenomena happening above continuum level, like acoustic glories, appear not to have a significant effect on the phases.
However there is a strong dependence on sunspot size, line-of-sight magnetic flux, and intensity contrast. As previously suggested by other groups, the region of wave scattering appears both horizontally smaller and vertically less extended than the region of acoustic power suppression, and occurs closer to the surface. Results presented here should help refine theoretical models of acoustic wave scattering by magnetic fields. Ingoing-ingoing phase differences between two measurement heights are also computed and show a complex pattern highly dependent on atmospheric height. In particular, a significant sensitivity of AIA signals to lower chromospheric layers is shown. Finally, ingoing-ingoing minus outgoing-outgoing phase differences between various measurement heights are discussed.
\end{abstract}


\end{opening}

\section{Introduction}

This article continues the study started in Couvidat (2013) based on data from the {\it Helioseismic and Magnetic Imager} instrument (HMI; Schou {\it et al.}, 2012) and the {\it Atmospheric Imaging Assembly} instrument (AIA; Lemen {\it et al.}, 2012) onboard the {\it Solar Dynamics Observatory} satellite (SDO). HMI measures the motion of the solar photosphere and produces Line-Of-Sight (LOS) observables every 45 seconds like, amongst others, Doppler velocity and continuum intensity. AIA images the Sun in 10 UV and EUV channels. The channels at 1600 \AA\ and 1700 \AA\ are adapted to helioseismic studies ({\it e.g.} Hill et {\it al.}, 2011; Howe {\it et al.}, 2011; Howe {\it et al.}, 2012; Rajaguru {\it et al.}, 2012).
Combining these four HMI and AIA datasets makes it possible to sample the entire photosphere. Indeed, the approximate formation heights of their signals are $0$, $\approx 150$, $\approx 360$, and $\approx 430$ km.
Therefore SDO allows multi-height helioseismic studies over the entire solar disk with unprecedented resolution and temporal cadence. This feature adds a third dimension to helioseismic measurements that should prove useful to deal with issues of wave propagation and properties in the photosphere.

Here an Hankel-Fourier transform (Braun, Duvall, and LaBonte, 1987) is applied to 14 sunspots and four quiet-Sun regions observed by HMI and AIA.
While Couvidat (2013) studied the power reduction and enhancement in the f- and p-oscillation modes, we focus on scattering phase shifts between ingoing and outgoing solar oscillation modes at given atmospheric heights, and between ingoing (or outgoing) modes at two different heights.
Braun {\it et al.} (1992) and Braun (1995) measured the phases of ingoing and outgoing waves at a given height and showed that scattering by the magnetic field of sunspots produces an outgoing-ingoing phase difference. Acoustic scattering at a given temporal frequency ({\it i.e.} under the assumption that any scattering inhomogeneity does not vary over the timescale considered) is usually described by a scattering matrix. The arguments of this matrix diagonal elements are the phase shifts resulting, at same wavelength and same frequency, from the scattering. Here we derive these shifts and neglect the matrix off-diagonal elements, describing any coupling between modes of different wavelengths.

 The mechanism responsible for scattering phase shifts was unclear at the time of Braun {\it et al.} (1992). However the authors mentioned that waves interacting with a simple uniform cylinder of increased wave speed produce a positive outgoing-ingoing phase shifts (their original paper claimed a reduced wave speed, but this was corrected in Braun, 1995). They concluded that no absorption mechanism of wave power is required. 
In other words, acoustic power reduction and wave scattering could be produced by two separate phenomena.
Note that the authors were careful to refer to wave speed and not sound speed: Both are quite different in the presence of strong magnetic fields (wave speed is a combination of sound and Alfv\'en speeds). Braun (1995) presented a more elaborate theory with a cylinder of increased wave-speed inhomogeneity of finite depth: He concluded that the scattering perturbation is most likely close to the solar surface. Fan, Braun, and Chou (1995) modeled the scattering of p-modes in a polytropic atmosphere by inhomogeneities in wave speed, density, and pressure. They did not explicitely account for the presence of magnetic field, but assumed an increase in wave-speed as a consequence of magnetic pressure inside the sunspots. They confirmed that the inhomogeneity characteristic depth plays a crucial role in determining the behaviour of the phase shifts. Inside a sunspot, the sound speed is reduced compared to the external medium and this translates into a negative outgoing-ingoing phase difference for acoustic waves ({\it e.g.} Gordovskyy and Jain, 2007a). However, in the presence of fields the wave speed is of the order of the fast magneto-acoustic speed, hence the assumption of Fan, Braun, and Chou (1995) of an increase in wave velocity in strong fields.

Acoustic power reduction in sunspots is believed to arise mainly from mode conversion ({\it e.g.} Cally, Crouch, and Braun, 2003): Incoming acoustic waves are partly converted into slow magneto-acoustic gravity (MAG) waves which guide the energy away from the surface. This energy appears lost to local helioseismology measurements. Most of the incoming waves are converted into fast MAG waves that will constitute the outgoing wavefield measured by Hankel transform. With this model Cally, Crouch, and Braun (2003) fitted the observed phase shifts of Braun (1995) with only magnetic fields (no thermal differences between the media outside and inside a sunspot were taken into account): It was concluded that the increase in p-mode phase speed inside a sunspot is due to magneto-acoustic coupling between acoustic waves and fast MAG waves, and that there is no need for a thermal perturbation to explain the results of Braun (1995). Crouch {\it et al.} (2005) later refined this model by adding the thermal perturbation resulting from the sunspot's magnetic field. The authors concluded that highly inclined fields ({\it i.e.} mainly in the sunspot penumbra) produce large phase differences, even though the main contribution to scattering phase shift seems to arise from the umbral core of sunspots. The indirect effect of magnetic field, {\it e.g.} the sound-speed reduction, was found to have a negligible impact. To summarize, the scattering phase shift results from an increased phase speed as the incoming acoustic waves are converted into fast MAG waves, while the acoustic power reduction results from part of the incoming waves being converted into slow MAG waves: power reduction and wave scattering originate from the same phenomenon. 

Fan, Braun, and Chou (1995) took a phenomenological approach based on Chou and Chen (1993) but neglected the Lorentz force. Gordovskyy and Jain (2007a) went a step further: They took into account the changes in pressure, density, and sound speed caused by the field, and the impact of Lorentz forces. They obtained that in sunspot models with vertical fields only, the reduced sound speed results in a negative scattering phase shift, but the Lorentz force produces an overall positive phase shift as the incoming acoustic waves turn into fast MAG modes. Previously, Gordovskyy, Jain, and Thompson (2006) had tested the impact of flaring magnetic fields, {\it i.e.} the presence of inclined fields in the penumbra of sunspots. However, they had not included the Lorentz force in their hydrodynamic equations. The result is that inclined-enough fields, unlike vertical fields, produce positive scattering shifts because they increase the wave speed. Finally, Gordovskyy and Jain (2007b) added to their model both inclined fields and the Lorentz force, resulting in positive phase shifts for all parameters tested.

Overall, a consensus has emerged that positive outgoing-ingoing phase shifts measured by Hankel transform with sunspot data result from strong magnetic fields mainly through mode conversion producing an increase in wave speed (despite a decrease in sound speed), and that the presence of inclined fields enhances this process. Furthermore wave scattering is thought to occur very close to the solar surface.

Measuring phase shifts is less straightforward than power reduction: Longer time series are required, and for the outgoing-ingoing shifts a correction of spurious values resulting from the presence of many unresolved modes is necessary. The phase differences retrieved are noisy and quite sensitive to this correction. Maybe because of these difficulties, only one dataset seems to have been published so far about such scattering phase shifts: The one of Braun {\it et al.} (1992) and Braun (1995), based on two sunspots. It seems worthwhile to revisit and expand their results by using more recent seismic data (from higher-quality instruments), and by increasing the sample size of sunspots (from two to 14). Moreover working with measurements at different photospheric heights instead of just one should provide additional information regarding the physics at play. Finally we also aim at providing some diagnostic plots to help refine and test the different models of wave scattering mentioned in this section.
  With a single measurement height, only the phase difference between radially ingoing and outgoing waves can be calculated. Combining two different heights allows the derivation of cross-spectra and phase differences for ingoing (or outgoing) waves between two separate altitudes. No correction is necessary in this case and this provides additional information about wave properties in the photosphere.

In Section 2 we briefly remind the reader of the Hankel-Fourier analysis. In Section 3 we describe the data used. In Sections 4 and 5 we present our results, and we conclude in Section 6.

\section{Hankel-Fourier Transform}

The Hankel-Fourier decomposition is extensively described in Braun, Duvall, and LaBonte (1987), Braun {\it et al.} (1992), and Braun (1995). Our implementation is summarized in Couvidat (2013). In polar coordinates [$(r,\theta)$] and as a function of time [$t$], the wave field [$\Phi$] is decomposed into the components
\begin{equation}
\Phi_{m;k}(r,\theta,t) = e^{i(m \theta + \omega t)} [ A_m(k,\omega) H^{(1)}_m(kr) + B_m(k,\omega) H^{(2)}_m(kr)]
\end{equation}
where $m$ is the azimuthal order, $k$ is the horizontal wavenumber, $\omega=2 \pi \nu$ is the cyclical frequency ($\nu$ is the temporal frequency), and $H^{(1)}_m$ and $H^{(2)}_m$ are Hankel functions of the first and second kind. The coefficients $A_m$ and $B_m$ are complex amplitudes of radially ingoing and radially outgoing waves, respectively. We compute these coefficients at four heights in the photosphere and pair them to compute cross-spectra. For instance, the cross-spectrum $AA_{m;ij}(k,\omega)$ of ingoing waves between the heights $i$ and $j$ for the azimuthal order $m$ is defined as 
\begin{equation}
AA_{m;i,j} = A_{m;i}^{*} \times A_{m;j}
\end{equation}
where $^{*}$ denotes the complex conjugate. The phase $\phi^{AA}_{m;ij}(k,\omega)$ of this cross-spectrum is the argument of $AA_{m;i,j}(k,\omega)$
\begin{equation}
\phi^{AA}_{m;ij}(k,\omega) = \mathrm{arctan}(\mathrm{Im}(AA_{m;i,j}(k,\omega))/\mathrm{Re}(AA_{m;i,j}(k,\omega)))
\end{equation}
where Im is the imaginary part and Re is the real part of a complex number. $\phi^{AA}_{m;ij}(k,\omega)$ is therefore the difference between the phase of $A_{m;j}(k,\omega)$ and the phase of $A_{m;i}(k,\omega)$. Similarly, $\phi^{AB}_{m;ij}(k,\omega)$ is the phase difference between outgoing and ingoing waves.
To increase the signal-to-noise (S/N) ratio on such phase measurements, an ``$m$-averaged'' phase $\phi^{AA}_{ij}(k,\omega)$ can be calculated, following Braun (1995)
\begin{equation}
\phi^{AA}_{ij}(k,\omega) = \mathrm{arg} \biggl( \frac{\sum_m \, W^{AA}_{m;ij}(k,\omega) \, \mathrm{exp}(i \, \phi^{AA}_{m;ij}(k,\omega))}{\sum_m \, W^{AA}_{m;ij}(k,\omega)} \biggr) 
\end{equation}
where arg is the argument and $W^{AA}_{m;ij}(k,\omega)$ is a weight calculated as
\begin{equation}
W^{AA}_{m;ij}(k,\omega) = |A_{m;i}(k,\omega)||A_{m;j}(k,\omega)|
\end{equation} 

The cross-spectrum $AB_{ij}(k,\omega)$ is used to measure the outgoing-ingoing phase shift at a given ($i=j$) or between two ($i \ne j$) atmospheric heights. However $\phi^{AB}_{ij}(k,\omega)$ must be corrected. The justification and details for this correction are presented in depth in Braun {\it et al.} (1992) and Braun (1995). Briefly, its necessity stems from the fact that within a grid element centered on ($k,\omega$) there are many unresolved oscillation modes. The correction [$\phi'_n(k,\omega)$] to apply is estimated from a set of solar oscillation frequencies, is independent of $m$, and is defined as 
\begin{equation}
\phi'_n(k,\omega) = -\biggl(\frac{\partial \omega}{\partial k}\biggr)^{-1} (\omega-\omega_n(k))
\end{equation}
where $\omega_n(k)$ is the angular frequency of the mode with radial order $n$ and wavenumber $k$.
Here we use a set of oscillation frequencies obtained with the MDI instrument (Scherrer {\it et al.}, 1995) onboard SOHO, during a dynamics run in 2007. This set was calculated and provided by \mbox{M.C.} Rabello-Soares and details are available in Rabello-Soares, Korzennik, and Schou (2008). For modes with orders $n$ from 0 to 9, the maximum degrees $\ell$ available in the set are: 890, 890, 860, 740, 530, 440, 370, 320, 260, and 220.
As pointed out in Braun (1995), the $\phi'_n(k,\omega)$ correction is quite sensitive to the accuracy of the oscillation frequencies used. On the other hand, phases of cross-spectra $\phi^{AA}_{ij}(k,\omega)$ and $\phi^{BB}_{ij}(k,\omega)$ (with $i \ne j$) do not require any correction.

To further increase the S/N ratio on these phase shifts, a ``ridge-averaged $m$-averaged'' phase difference can be calculated, for instance $\phi^{AB}_{ij}(k,n)$, again following Braun (1995)
\begin{equation}
\phi^{AB}_{ij}(k,n) = \mathrm{arg} \biggl( \frac{\sum_m \int_{\omega_n(k)-\delta\omega}^{\omega_n(k)+\delta\omega} \, W^{AB}_{m;ij}(k,\omega) \, \mathrm{exp}(i[\phi^{AB}_{m;ij}(k,\omega)-\phi'_n(k,\omega)]) \, d\omega}{\sum_m \int_{\omega_n(k)-\delta\omega}^{\omega_n(k)+\delta\omega} \, W^{AB}_{m;ij}(k,\omega) \, d\omega} \biggr)
\end{equation}
where the frequency range $[\omega_n(k)-\delta\omega,\omega_n(k)+\delta\omega]$ is selected to span the $n$ ridge width. The denominators in Equations 4 and 7 can be dropped since we take the argument.
A mask is applied to the cross-spectra, to only select power within this ridge.
Braun (1995) and Fan, Braun, and Chou (1995) demonstrated that $\phi^{AB}_{m;ii}(k,\omega)$ peaks at the azimuthal order $m=0$ and drops quickly with $m$. Figure 3 of Fan, Braun, and Chou (1995) shows that for $|m|/k \approx 25$ Mm, tan($\phi^{AB}_{m;ii}(k,\omega)$) $\approx 0.2$ tan($\phi^{AB}_{0;ii}(k,\omega)$). Therefore, in the calculation of the ``$m$-averaged'' phase differences, we select as maximum value of $|m|$ at each $k$ the one such as $|m|/k$ is close to 25 Mm.

Even though we refer to $\phi^{AA}_{ij}(k,\omega)$ (Equation 4) and $\phi^{AB}_{ij}(k,n)$ (Equation 7) as ``average'' phase differences, they do not result from a straightforward averaging but are the phase differences of an average over the complex amplitudes $AA_{m;i,j}$ and $AB_{m;i,j}$ (sometimes called a vector-average phase). Both averaging schemes can produce results that are quite different. Using the complex representation of phases gets rid of their $2\pi$ discontinuity, and, even though this is not a strong scientific justification, is the averaging scheme utilized in Braun (1995) which is used as the comparison basis for our results and for numerous papers based on his results.
However, phase-difference algebraic average and vector average can significantly differ: they both qualitatively return (mostly) positive phase shifts in sunspots but they quantitatively differ, especially at high $\ell$.

Moreover, it must be pointed out that only the phases of individual modes make physical sense, but the high noise level on such individual measurements prevents their use and forces us to resort to some averaging scheme. Only if the phase shifts of individual modes are independent of $m$ does the averaged phase shift represent the shift of an individual mode, but because the shifts of individual modes strongly depend on $m$ the physical meaning of the ``average'' shifts computed by Equations 4 and 7 is unclear: they do not represent the phase shift of an individual mode, although they are the phase shift of some kind of averaged quantity. The reader should keep this fact in mind when interpreting our results.

Finally, we remind the reader that the polar coordinate system used is centered on a region of interest, {\it e.g.} the center of a sunspot, and the ingoing- and outgoing-wave decomposition is calculated in an annulus with inner radius $R_\mathrm{min}$ selected to exclude this sunspot.

\section{Data Used}

The same datasets are used as in Couvidat (2013): HMI continuum intensity, HMI LOS Doppler velocity, AIA 1700 \AA, and AIA 1600 \AA\ intensity maps. The LOS magnetic fluxes are estimated from HMI LOS magnetic-field strength. Each datacube is tracked and Postel-projected  for four days at Snodgrass rotation rate using the mtrack software.
The spatial resolution is $dx=0.09$ heliocentric degrees and the temporal cadence is $dt=45$ seconds.
Tracked cubes have the format: $384 \times 384 \times 7681$, where the first two dimensions are the spatial coordinates, and the last one is time.
Only $62.5$ hours are selected from each cube.
We set  $R_\mathrm{min}=25.15$ Mm and $R_\mathrm{max}=209$ Mm (annulus outer radius).
The resolution in wavenumber [$k$] is $dk=2\pi/(R_\mathrm{max}-R_\mathrm{min})=0.0342$ Mm$^{-1}$, {\it i.e.} a resolution in angular degree [$\ell$] equal to $d\ell=dk \times R_{\odot}=23.8$ (where $R_{\odot}=696$ Mm is the solar radius).
The resolution in temporal frequency [$\nu$] is $d\nu=4.44$ $\mu$Hz.
Table 1 of Couvidat (2013) lists all of the sunspots studied: note that active region NOAA 11263 is not used here because it is too asymmetrical.

\section{Results: Outgoing-Ingoing Scattering Phase Shifts}

Braun {\it et al.} (1992) and Braun (1995) pionereed the calculation of scattering phase shifts $\phi^{AB}_{ii}(k,n)$ between outgoing and ingoing waves at the same atmospheric height, using the Hankel transform. Here we expand on their results by studying more sunspots and by computing $\phi^{AB}_{ii}(k,n)$ at four different measurement heights. First, for comparison purpose, we present our results as a function of angular degree $\ell$ and frequency $\nu$. Again, we use a vector-average of phase differences (like Braun, 1995) rather than a simple average.

\begin{figure}
\centering
\includegraphics[width=\textwidth]{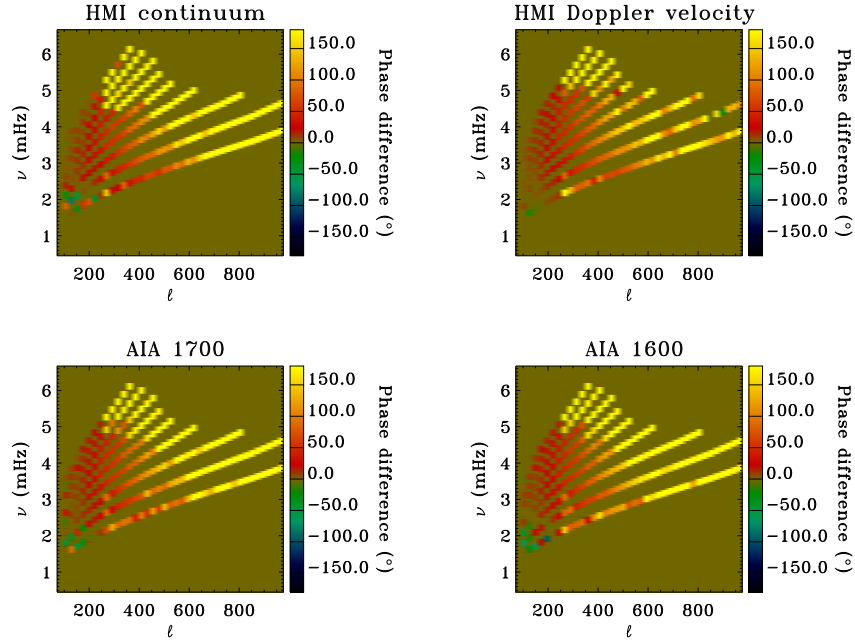}
\caption{Phase difference $\phi^{AB}_{ii}(k,n)$ between outgoing and ingoing waves at four different atmospheric heights. Each map is an average over the results of the six strongest sunspots of our sample. The f-mode ridge is not shown as no phase difference could be measured.}
\label{firstfig}       
\end{figure}

Figure \ref{firstfig} shows 2D  maps of $\phi^{AB}_{ii}(k,n)$ at the four measurement heights. The phase shifts are averaged over the six strongest --- in terms of LOS magnetic flux --- sunspots studied: NOAA 11092, 11289, 11314, 11384, 11408, and 11410. The four maps look very similar: Each shows an increase in phase difference toward larger $\ell$ and larger $\nu$. No f-modes are shown because no phase shift could be measured for $n=0$. Our results for the strongest sunspots are very similar to those of Braun (1995). Moreover, the phase-difference maps for quiet-Sun regions show a relatively small negative $\phi^{AB}_{ii}(k,n)$ increasing (in absolute value) with $\ell$. Braun (1995) argued that this negative phase shift is due to an imperfect correction $\phi'_n(k,\omega)$, resulting from errors in the frequencies and/or angular degrees of the oscillation modes measured. Indeed, a simple reduction in $\ell$ by a couple of parts in a thousand flattens the quiet-Sun curves of $\phi^{AB}_{ii}(k,n)$ as a function of $\ell$. 
Following Braun {\it et al.} (1992), in the rest of this paper we subtract an average over the four quiet-Sun regions of the phase shifts from all of the shifts obtained in sunspots: We are primarily interested in separating the impact of magnetic fields of sunspots from other influences. 

The shifts are only known modulo $2\pi$, which means that some values can be corrected by adding or subtracting $360^{\circ}$. Although there is a part of subjectivity in selecting which values to correct, due to their noisy nature, in most cases detecting ``bad'' phases is straightforward: {\it e.g.}, if a shift at a given $\ell$ is strongly negative while the surrounding (in terms of $\ell$) shifts are strongly positive, $360^{\circ}$ is added to the negative phase shift.

\begin{figure}
\centering
\includegraphics[width=\textwidth]{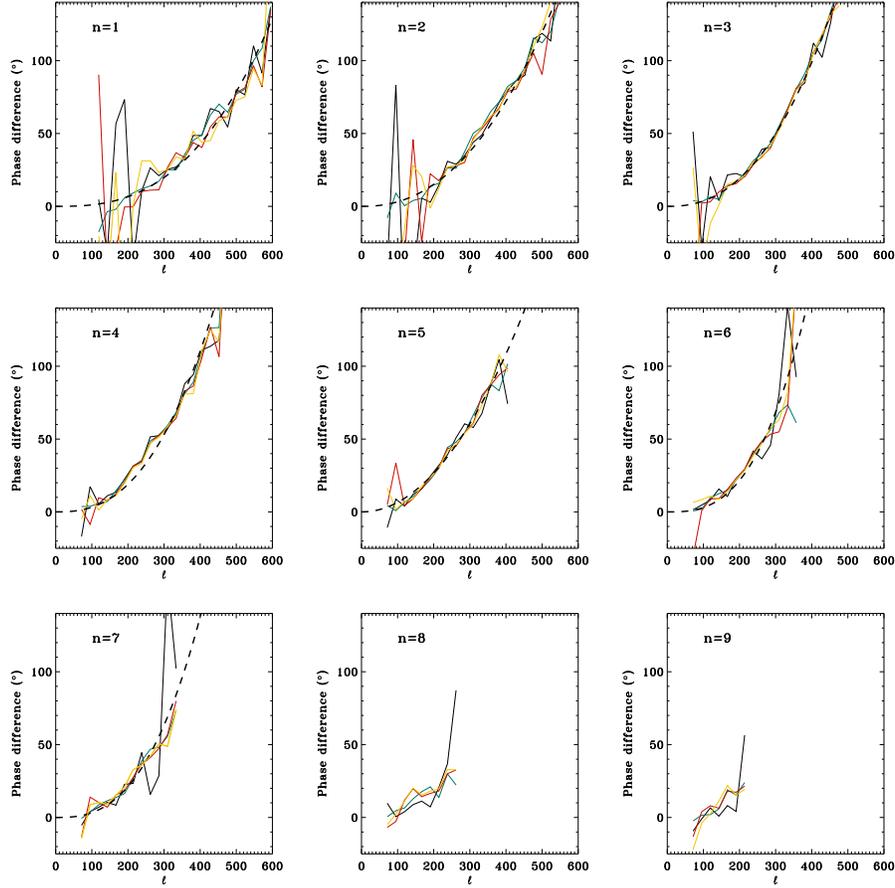}
\caption{Phase difference $\phi^{AB}_{ii}(k,n)$, averaged over the six strongest sunspots, as a function of angular degree $\ell$ for p-modes with radial order $n=1$ (upper left panel) to $n=9$ (lower right panel). The solid black lines are for the HMI continuum data, the green lines are for the HMI Doppler velocity data, the red lines are for the AIA 1700 \AA\ data, and the yellow lines are for the AIA 1600 \AA\ data. The dashed black lines are the result of a fit by a power law.}
\label{thirdfig}       
\end{figure}

Figure \ref{thirdfig} shows the average of $\phi^{AB}_{ii}(k,n)$ over the six strongest sunspots as a function of angular degree $\ell$ for p-modes with radial order $n=1$ (upper left panel) to $n=9$ (lower right panel) and for the four measurement heights. At low-$n$ we display phase shifts for $\ell$ larger than those shown in Braun (1995). The limiting factor regarding which $\ell$ (or $\nu$) values can be reached is mainly the set of solar oscillation frequencies used to determine the $\phi'_n(k,\omega)$ correction.

A new result is that the outgoing-ingoing phase shifts appear mostly independent of the atmospheric height at which the signal is measured. Indeed, for p-modes $n=1$ to $n=5$, the shifts do not seem to change with height (again, after a quiet-Sun average was subtracted). This lack of strong dependence on atmospheric height indicates that most of the scattering of acoustic waves occurs very close to, or below, the continuum level and that not much occurs higher in the photosphere. This situation contrasts significantly with what was obtained for the power-reduction of outgoing waves in sunspots (Couvidat 2013): A strong dependence on measurement height was observed. In particular acoustic glories (haloes) that strongly impact the acoustic power reduction coefficients seem to have less influence on the scattering phase shifts. Acoustic glories are most obvious in the $4.5-5.5$ mHz range, where our ability to measure phase shifts is limited (the noise level is higher), but power enhancement appears at frequencies as low as 4 mHz in a few sunspots while in more typical active regions and at $\nu=4$ mHz the power reduction coefficient drops from $\approx 0.2$ at continuum level to less than $0.1$ at AIA 1600 level. Here, for $n=4$ to 7 the scattering phases at $\nu=4-4.5$ mHz do not show such a significant change with measurement height. At higher $n$ the HMI continuum might differ from the other datasets, but the results are too noisy to draw firm conclusions.  
Gordovskyy and Jain (2007b) referred to a preliminary work suggesting that a significant contribution to the phase shifts arise from the part of the flux tube above the surface. However, our results do not support this statement and hint at a limited role in the scattering for the tube above the surface. 

Following Gordovskyy and Jain (2007b) we fitted $\mathrm{tan}(\phi^{AB}_{ii}(k,n))$ as a function of $\ell$ for a given $n$ by a power law ($\mathrm{tan}(\phi^{AB}_{ii}) \propto \ell^q$), but only in a restricted range of $\ell$ and for modes $n=1$ to $n=6$. The best fits return $3.38 \le q \le 4.8$, compatible with a strong convergence of the magnetic field: In the model of Gordovskyy and Jain (2007b), convergence refers to how quickly the flux tube cross-section decreases by a factor of two with depth. In other words, the measured phase shifts are compatible with the presence of strongly inclined fields in the sunspots studied. We also fitted $\phi^{AB}_{ii}(k,n)$ as a function of $\ell$ by a power law ($\phi^{AB}_{ii} \propto \ell^q$): A better agreement is obtained over the entire range of accessible $\ell$, with $2 \le q \le 2.8$ (for $n=1$ to $n=6$). 
This power law does not match Figure 8 of Fan, Braun, and Chou (1995) and Figure 9 of Gordovskyy and Jain (2007b): On these two figures the derivative of $\phi^{AB}_{ii}(k,n)$ with respect to $\ell$ decreases for, roughly, $\ell \ge 300$, which is clearly not the case on Figure \ref{thirdfig}. Fan, Braun, and Chou (1995) concluded that the Born approximation they used does not hold for the large phase differences reached at higher $\ell$. 

\begin{figure}
\centering
\includegraphics[width=\textwidth]{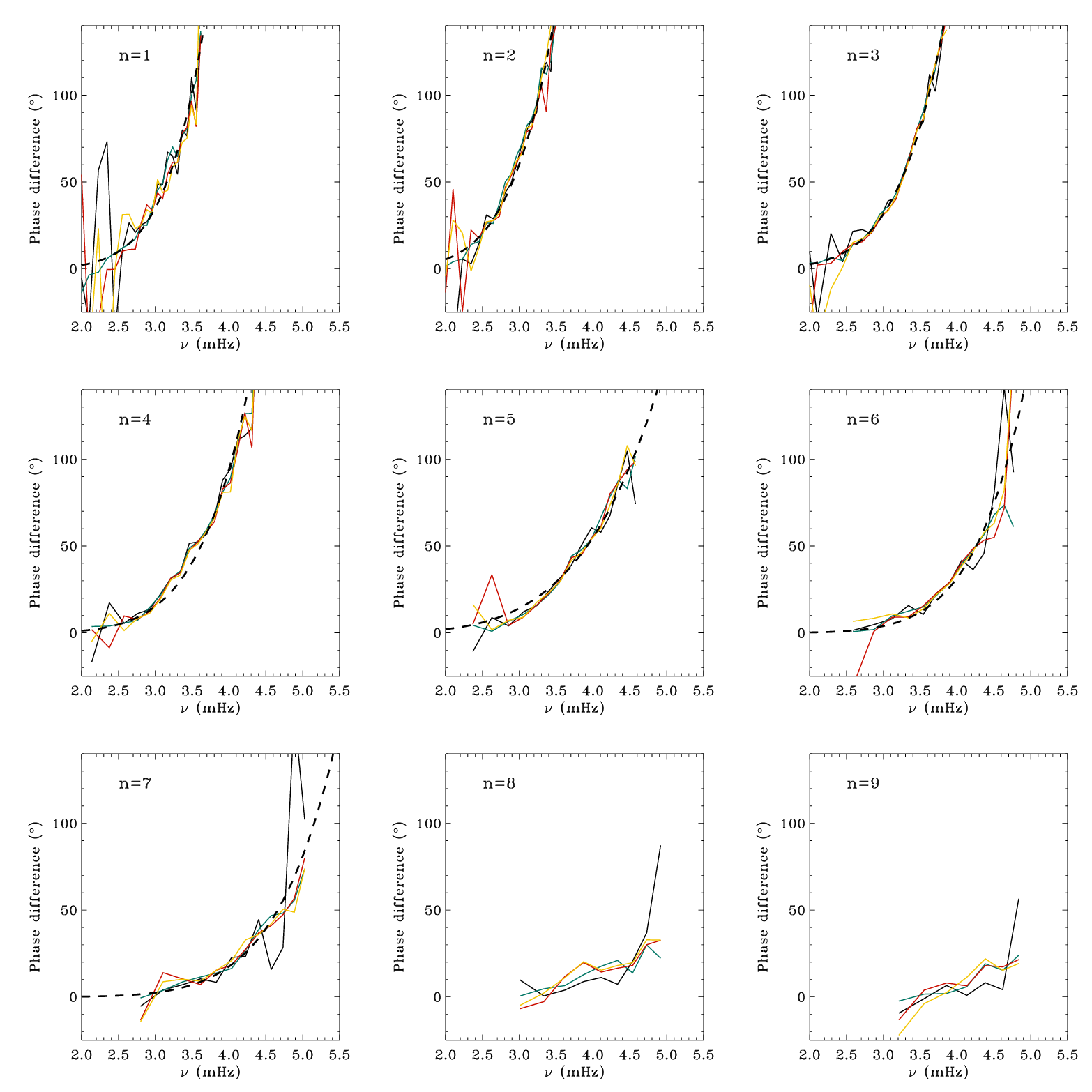}
\caption{Phase difference $\phi^{AB}_{ii}(k,n)$, averaged over the six strongest sunspots, as a function of frequency $\nu$ for p-modes with radial order $n=1$ (upper left panel) to $n=9$ (lower right panel). The solid black lines are for the HMI continuum data, the green lines are for the HMI Doppler velocity data, the red lines are for the AIA 1700 \AA\ data, and the yellow lines are for the AIA 1600 \AA\ data. The dashed black lines are the result of a fit by a power law.}
\label{thirdfigc}       
\end{figure}

Figure \ref{thirdfigc} shows the average of $\phi^{AB}_{ii}(k,n)$ as a function of mode frequencies $\nu$ and for different radial orders $n$. Again, these results are very close to those obtained by Braun (1995) as shown on Figure 11 of Crouch {\it et al.} (2005), and the scattering phase shift clearly increases with $\nu$.
We also fitted  $\phi^{AB}_{ii}(k,n)$ as a function of $\nu$ by a power law ($\phi^{AB}_{ii} \propto \nu^q$) over the entire range of $\nu$ accessible and obtained values of $q$ ranging from $4.7$ to $6.9$ (for $n=1$ to $n=6$). 

The presence of negative phase shifts at low $\ell$ and $\nu$ ({\it e.g.} on Figures \ref{firstfig}, \ref{thirdfig}, and \ref{thirdfigc}) does not seem to be reflected in the models of Fan, Braun, and Chou (1995), Crouch {\it et al.} (2005), and Gordovskyy and Jain (2007b), but Braun (1995) also measured some negative shifts. Lets assume that these values are real and are not introduced by a faulty $\phi'_n(k,\omega)$ correction. Negative phase shifts were mentioned in Cally, Crouch, and Braun (2003) and interpreted as scattering occuring in the sunspot umbra where the field is nearly vertical. These authors also predicted that this would lead to a dip at low $m$ in $\phi^{AB}_{m;ii}$. As will be shown later, the evidence is rather weak here: Some oscillation modes show a dip, some do not, but the uncertainty on the measured $\phi^{AB}_{m;ii}$ is too large to draw any firm conclusion. Gordovskyy, Jain, and Thompson (2006) also obtained negative phase shifts when neglecting the Lorentz force and for low convergence ({\it i.e.} more vertical) fields, resulting from the reduced sound speed inside sunspots. Therefore both models link the existence of negative phase shifts to vertical fields. 

\begin{figure}
\centering
\includegraphics[width=\textwidth]{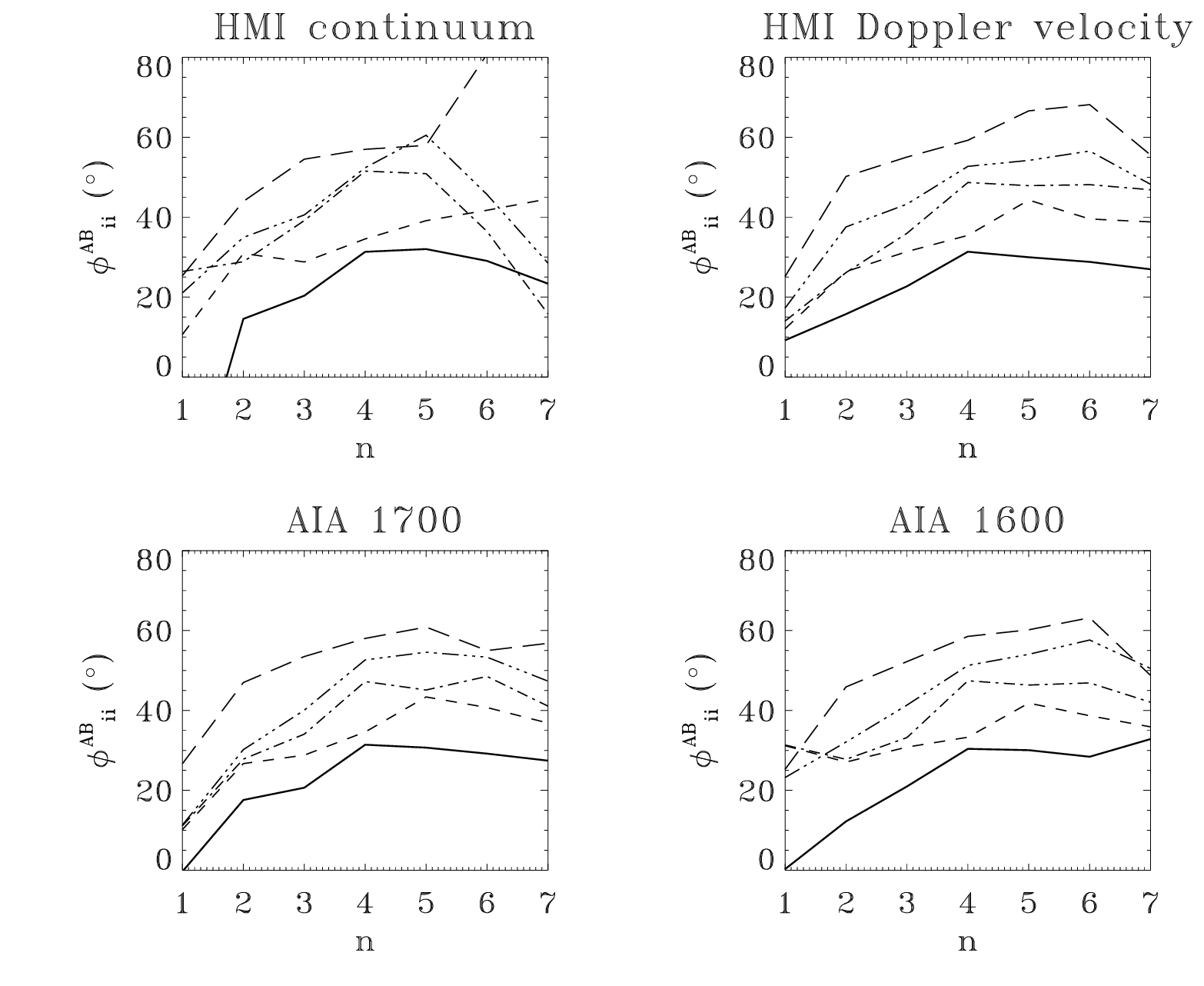}
\caption{Phase difference $\phi^{AB}_{ii}(k,n)$, averaged over the six strongest sunspots, as a function of the radial order $n$ of p-modes, and for the four different datasets. Thick solid lines are for modes with $\ell=214$, short-dashed lines are for $\ell=238$, dash-dotted lines are for $\ell=262$, dash-dot-dot-dotted lines are for $\ell=286$, and long-dashed lines are for $\ell=309$.}
\label{thirdfigb}       
\end{figure}

Figure \ref{thirdfigb} shows the phase shifts $\phi^{AB}_{ii}(k,n)$, averaged over the six strongest sunspots, as a function of radial order $n$ of p-modes and for different angular degrees $\ell$. Again, a lack of strong dependence on measurement height is visible.
Gordovskyy and Jain (2007b) mentioned that phase shifts vary roughly proportionally to $n$ based only on their low-$n$ results. On Figure \ref{thirdfigb} it appears that $\phi^{AB}_{ii}(k,n)$ increases with $n$ but only up to $n \approx 5$, and then decreases for higher $n$. This is in agreement with Fan {\it et al.} (1995) who interpreted this result as being due to a shallow inhomogeneity region.

\begin{figure}
\centering
\includegraphics[width=\textwidth]{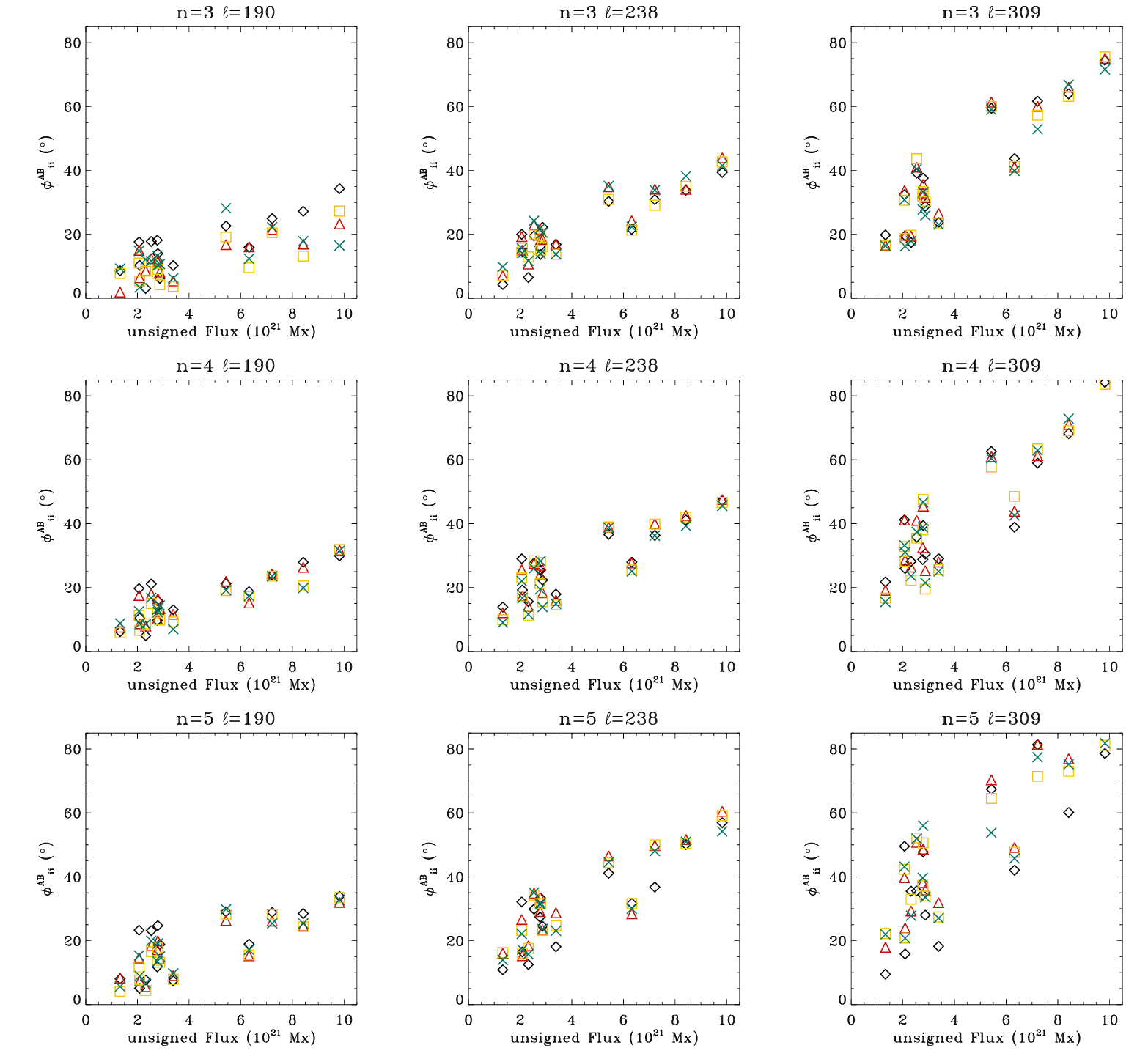}
\caption{Phase difference $\phi^{AB}_{ii}(k,n)$, averaged over the six strongest sunspots, as a function of the unsigned magnetic flux in the inner disk of radius $R_\mathrm{min}$, for radial orders $n=3$, $n=4$, and $n=5$, and three angular degrees $\ell=190$, $238$, and $309$. Black diamonds are for the HMI continuum data, red triangles are for the HMI Doppler velocity data, yellow squares are for the AIA 1700 \AA\ data, and green crosses are for the AIA 1600 \AA\ data.}
\label{fifthfigb}       
\end{figure}

Figure \ref{fifthfigb} shows $\phi^{AB}_{ii}(k,n)$ as a function of LOS magnetic flux in the inner disk of radius $R_\mathrm{min}$, and for p-modes with $n=3$, 4, and 5, and three different $\ell$. There is a strong dependence on LOS magnetic flux, with larger fluxes producing larger shifts. This kind of plots should prove useful to better constrain the parameters of the different wave scattering models.
If we fit $\phi^{AB}_{ii}(k,n)$ as a function of the peak LOS field strength $B_0$ in sunspots by a power law ($\phi^{AB}_{ii} \propto B_0^q$ where $B_0$ comes from HMI magnetograms), the best fits for modes $n=3$ and $4$ return values $q \le 1.5$ (at $\ell=309$). Gordovskyy and Jain (2007a) concluded that the outgoing-ingoing phase shifts are proportional to $B^2$, where $B$ is the characteristic magnetic field strength in their model which only includes vertical fields, unlike actual sunspots: Perhaps it is not surprising that our fit favors $q <2$, compatible with the notion that a significant part of wave scattering occurs in the inclined fields of the penumbra.

\begin{figure}
\centering
\includegraphics[width=\textwidth]{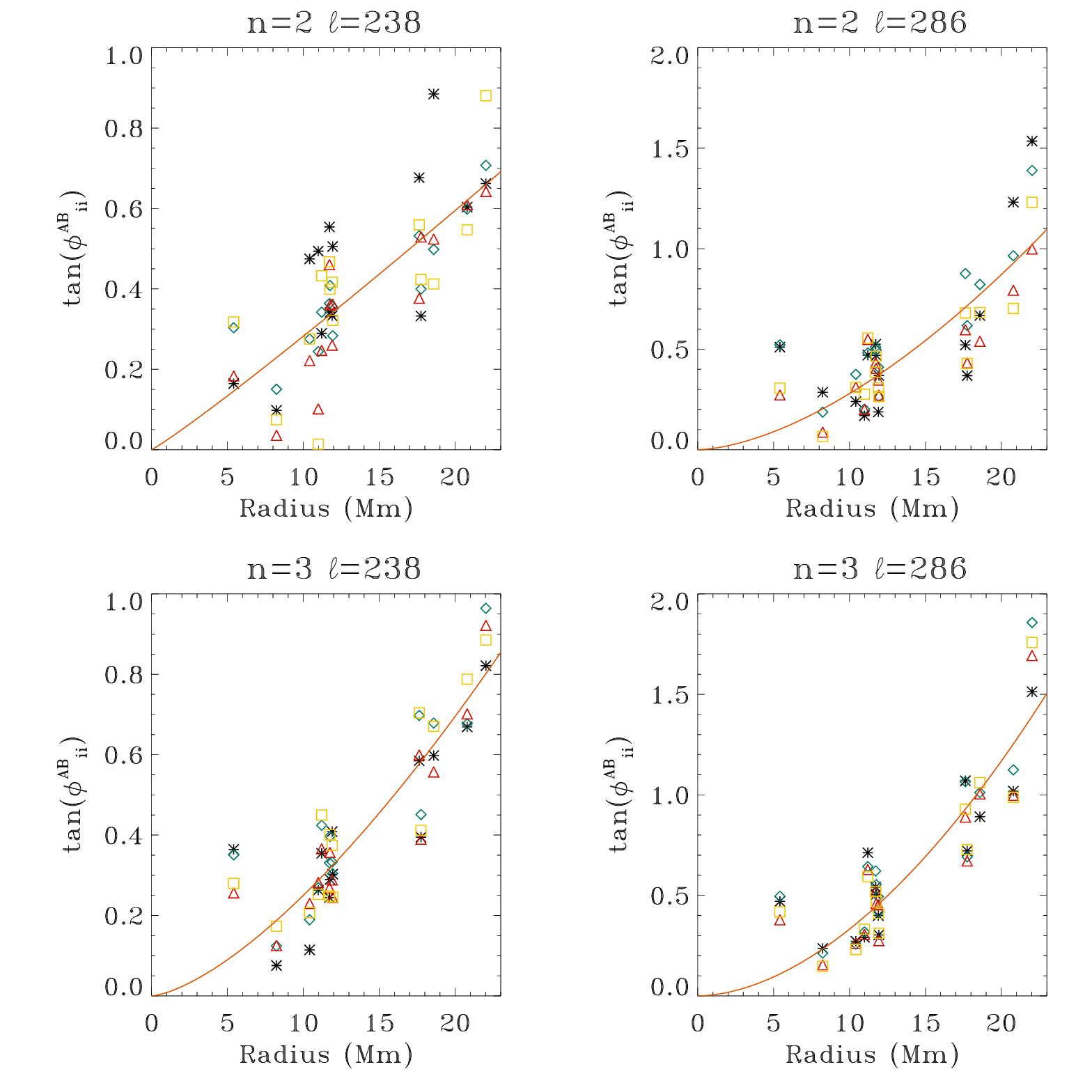}
\caption{Tangent of the average phase difference $\phi^{AB}_{ii}(k,n)$ for $n=2$ and $n=3$ in the six strongest sunspots, as a function of the sunspot radius, and for the four atmospheric heights. Black stars are for the HMI continuum, green diamonds are for the HMI Doppler velocity, red triangles are for the AIA 1700 \AA\ data, and yellow squares are for the AIA 1600 \AA\ data. The solid red line is the result of a fit by a power law. Left panel is for $\ell=238$, right panel is for $\ell=286$.}
\label{sixthfigb}       
\end{figure}

Similarly Figure \ref{sixthfigb} shows the tangent of the average of $\phi^{AB}_{ii}(k,n)$ over the six strongest sunspots and for $n=2$ and $n=3$, as a function of sunspot radius $R$ measured in continuum data ($R=\sqrt{S/\pi}$ where $S$ is the surface area of sunspots), and for the four measurement heights. A pixel is considered part of the sunspot if its continuum intensity is lower than 95\% of the average intensity outside the disk of radius $R_\mathrm{min}$. Unsurprisingly larger sunspots produce larger phase shifts. This was expected as $S$ and LOS magnetic flux are highly correlated. The increase in $\mathrm{tan}(\phi^{AB}_{ii}(k,n))$ with $R$ can be compared to the theoretical results of Gordovskyy and Jain (2007b): They suggested that for weakly converging fields $\mathrm{tan}(\phi^{AB}_{ii}(k,n))$ varies as $R^q$ where $q \approx 1$, for $k \ge 1/R$. Here, we plotted the results for two values of $\ell$ (238 and 286), so that $k > 1/R$ for all $R$. Our best fits return $q=1.47$ at $\ell=238$ and $q=1.81$ at $\ell=286$: These values are larger than expected by Gordovskyy and Jain (2007b) and favor a strong magnetic-field convergence in the sunspots studied.

\begin{figure}
\centering
\includegraphics[width=\textwidth]{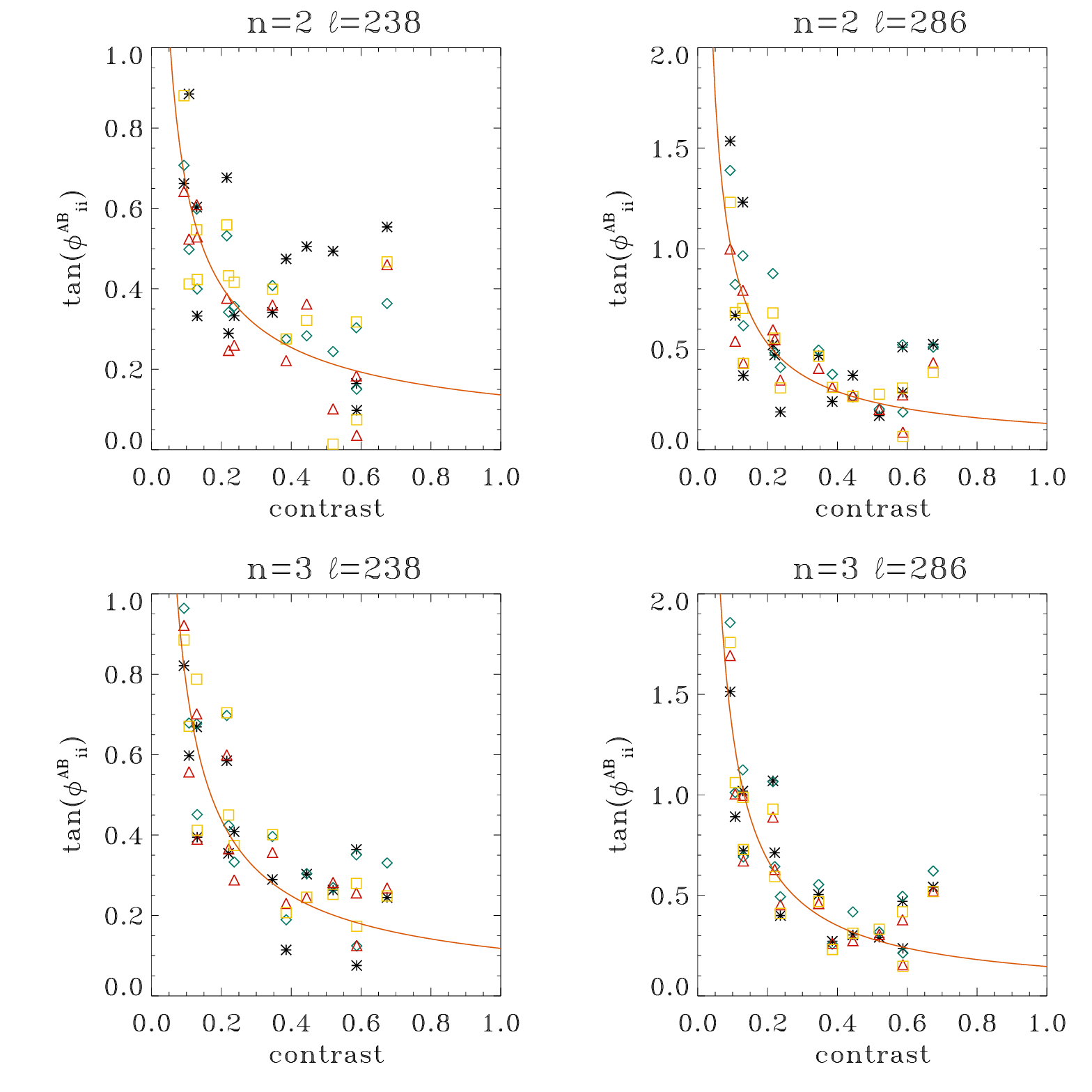}
\caption{Tangent of the average phase difference $\phi^{AB}_{ii}(k,n)$ over the six strongest sunspots, as a function of the contrast of sunspots, and for two radial orders $n=2$ (upper panels) and $n=3$ (lower panels). Black stars are for the HMI continuum, green diamonds are for the HMI Doppler velocity, red triangles are for the AIA 1700 \AA\ data, and yellow squares are for the AIA 1600 \AA\ data. The solid red line is the result of a fit by a power law (the data point contrast=1 and tangent=0 was added for the fit). Left panels are for $\ell=238$, right panels are for $\ell=286$.}
\label{fifthfigbb}       
\end{figure}

We also plotted the tangent of the average phase shift $\phi^{AB}_{ii}(k,n)$ for $n=2$ and $n=3$ over the six strongest sunspots, as a function of intensity contrast in sunspots (Figure \ref{fifthfigbb}). Intensity contrast is defined as the ratio of the lowest continuum intensity in sunspot umbra over the average intensity in a nearby quiet-Sun region. Overall phase shifts $\mathrm{tan}(\phi^{AB}_{ii}(k,n))$ increase when the contrasts decrease but there seems to be a plateau for contrasts higher than $\approx 0.4$ and the phase shift stays nearly constant in the range $0.4-0.7$.

\begin{figure}
\centering
\includegraphics[width=\textwidth]{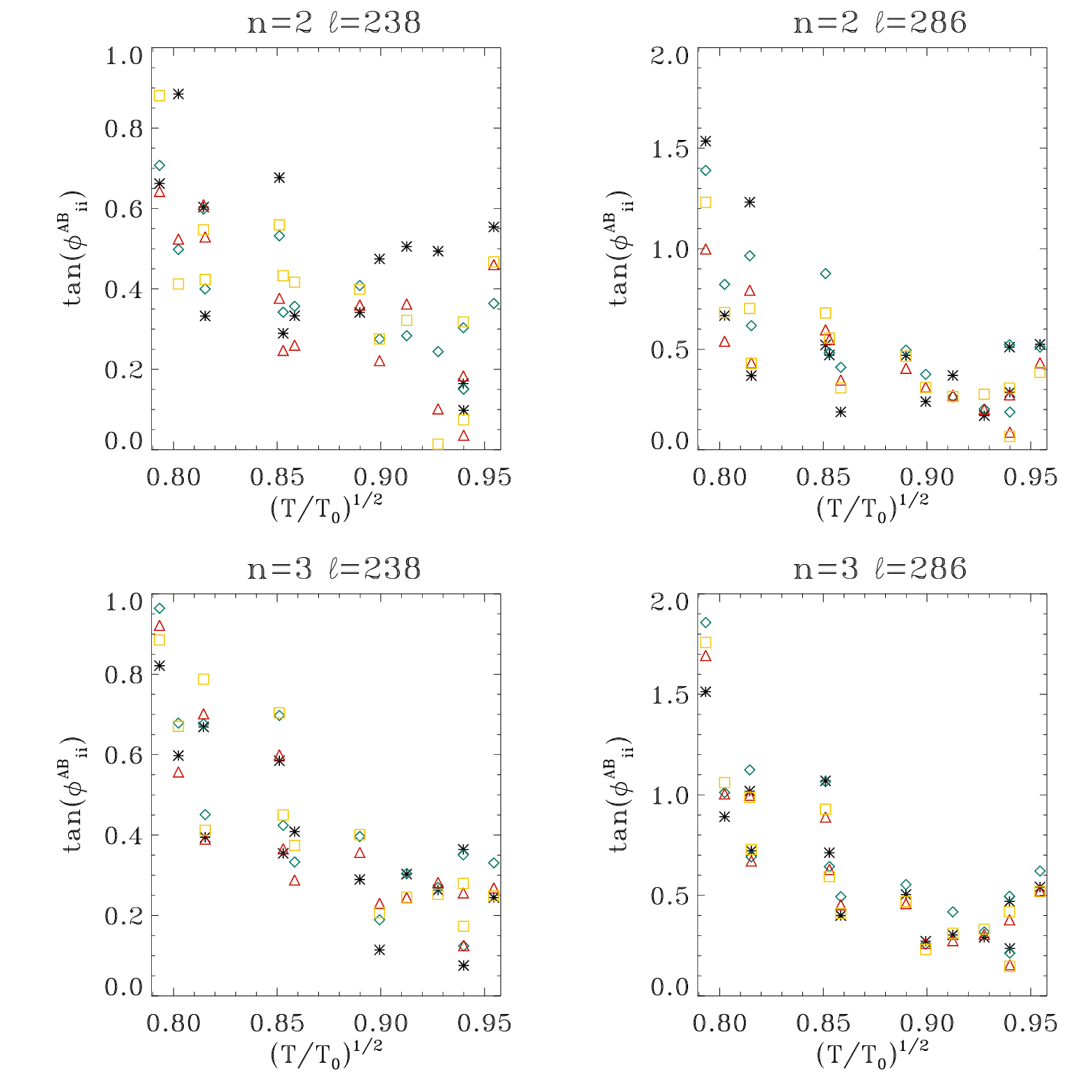}
\caption{Tangent of the average phase difference $\phi^{AB}_{ii}(k,n)$ over the six strongest sunspots, as a function of the square root of surface temperature $T$ of sunspots (divided by effective temperature $T_0=5777$ K) derived with the Eddington-Barbier approximation, and for two radial orders $n=2$ (upper panels) and $n=3$ (lower panels). Left panels are for $\ell=238$, right panels are for $\ell=286$. Black stars are for the HMI continuum, green diamonds are for the HMI Doppler velocity, red triangles are for the AIA 1700 \AA\ data, and yellow squares are for the AIA 1600 \AA\ data.}
\label{fifthfigbbb}       
\end{figure}

Using the Eddington-Barbier approximation ({\it e.g.} Borrero and Ichimoto, 2011) with a solar effective temperature of 5777 K at optical depth $\tau=2/3$, the contrasts are converted into surface temperatures $T$. Assuming that the perfect gas law is valid inside sunspots, the sound speed [$c_s$] is expected to vary as $c_s \propto T^{1/2}$ (at constant mean molecular weight). We plotted $\mathrm{tan}(\phi^{AB}_{ii}(k,n))$ as a function of $T^{1/2}$ on Figure \ref{fifthfigbbb}. Since $T^{1/2}$ is a proxy for $c_s$ at the solar surface, it appears that the scattering phase shifts increase when $c_s$ drops, but not in a linear fashion. For a sound speed in the range $0.9$ to $0.96$ times the quiet-Sun value, the scattering phases seem to be almost independent of $c_s$. Then, as $c_s$ further drops, there is a steep rise in $\mathrm{tan}(\phi^{AB}_{ii})$.

\begin{figure}
\centering
\includegraphics[width=\textwidth]{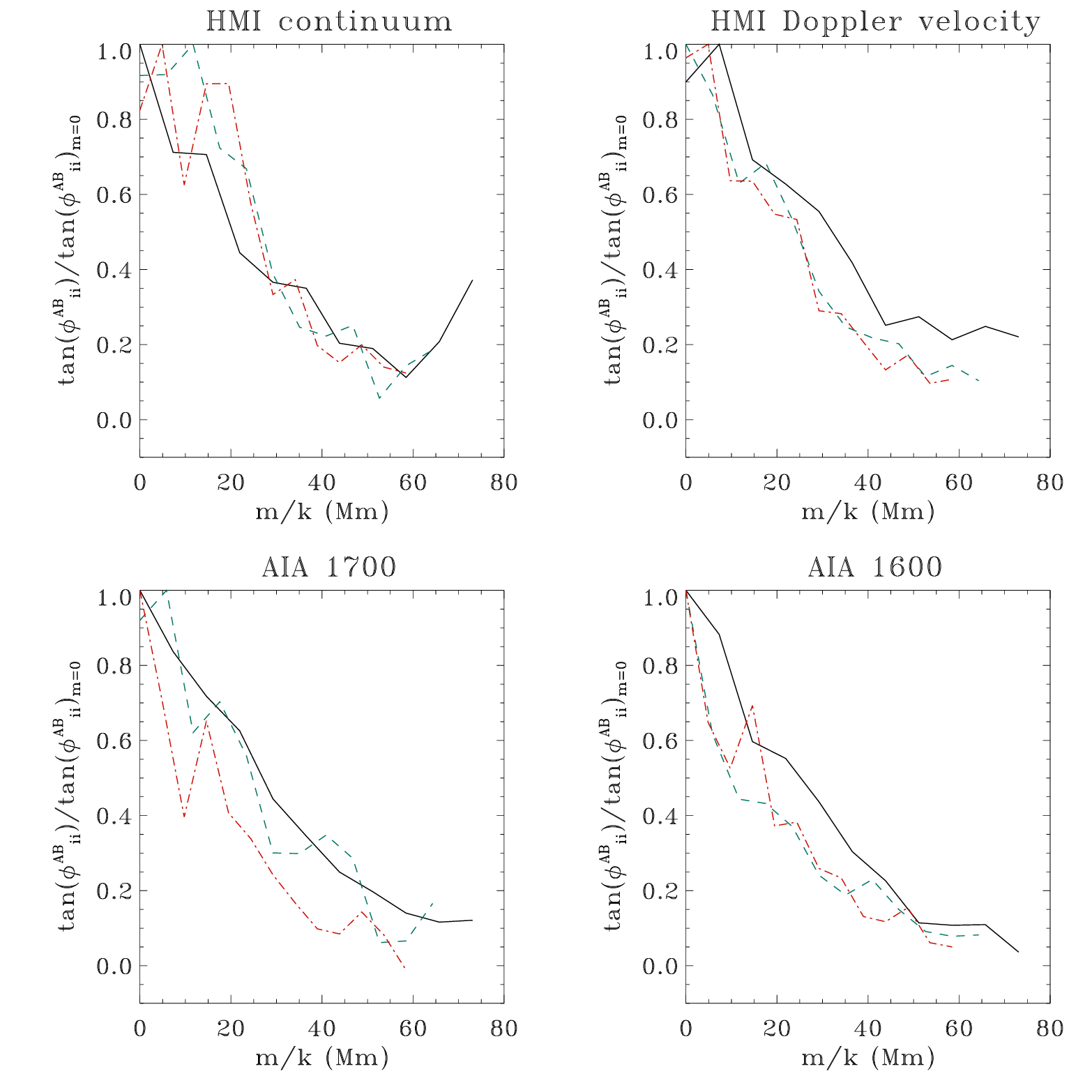}
\caption{Normalized tangent of the average of $\phi^{AB}_{m;ii}(k,n)$ over the six strongest sunspots as a function of the impact radius $m/k$, for three different $\ell$ and for radial order $n=3$, at the four atmospheric heights. Black solid line is for $\ell=262$, green dashed line is for $\ell=286$, and red dot-dashed line is for $\ell=309$.}
\label{sixthfigc}       
\end{figure}

Figure \ref{sixthfigc} shows the tangent of the average of $\phi^{AB}_{m;ii}(k,n)$ over the six strongest sunspots as a function of impact parameter $m/k$ and for radial order $n=3$, at the four measurement heights. As Braun (1995) first demonstrated, $\phi^{AB}_{m;ii}(k,n)$ quickly vanishes with $m$. This supports the idea that acoustic scattering occurs mostly inside the disk of radius $R_\mathrm{min}$. Moreover, this decrease with $m$ is steeper than the one measured with the power-reduction coefficients (see Figure 9 of Couvidat, 2013, for comparison). The phase difference is expected to drop to zero for values of $|m|/k$ larger than, or equal to, the radius of the scatterer. Based on Figure 3 of Fan, Braun, and Chou (1995), with the caveat that these authors assumed a Gaussian distribution for the scattering inhomogeneity and ignored any direct magnetic field effect, Figure \ref{sixthfigc} implies that for the six strongest sunspots studied the scatterer has a radius of $25-30$ Mm (roughly the value at which $\mathrm{tan}(\phi^{AB}_{m;ii})/\mathrm{tan}(\phi^{AB}_{0;ii}) = 0.4$): This is larger than the radius of any of these sunspots (a result also obtained by Fan, Braun, and Chou, 1995). Therefore, the scattering of acoustic waves extends horizontally beyond the visible radius of sunspots. We did not unambiguously observe the dip in phase shift that Cally, Crouch, and Braun (2003) expected at low impact parameter, resulting from the umbral core producing negative phase shifts (with their model): Some curves of $\mathrm{tan}(\phi^{AB}_{m;ii})$ as a function of $m$ show such a dip, while others do not.

\begin{figure}
\centering
\includegraphics[width=\textwidth]{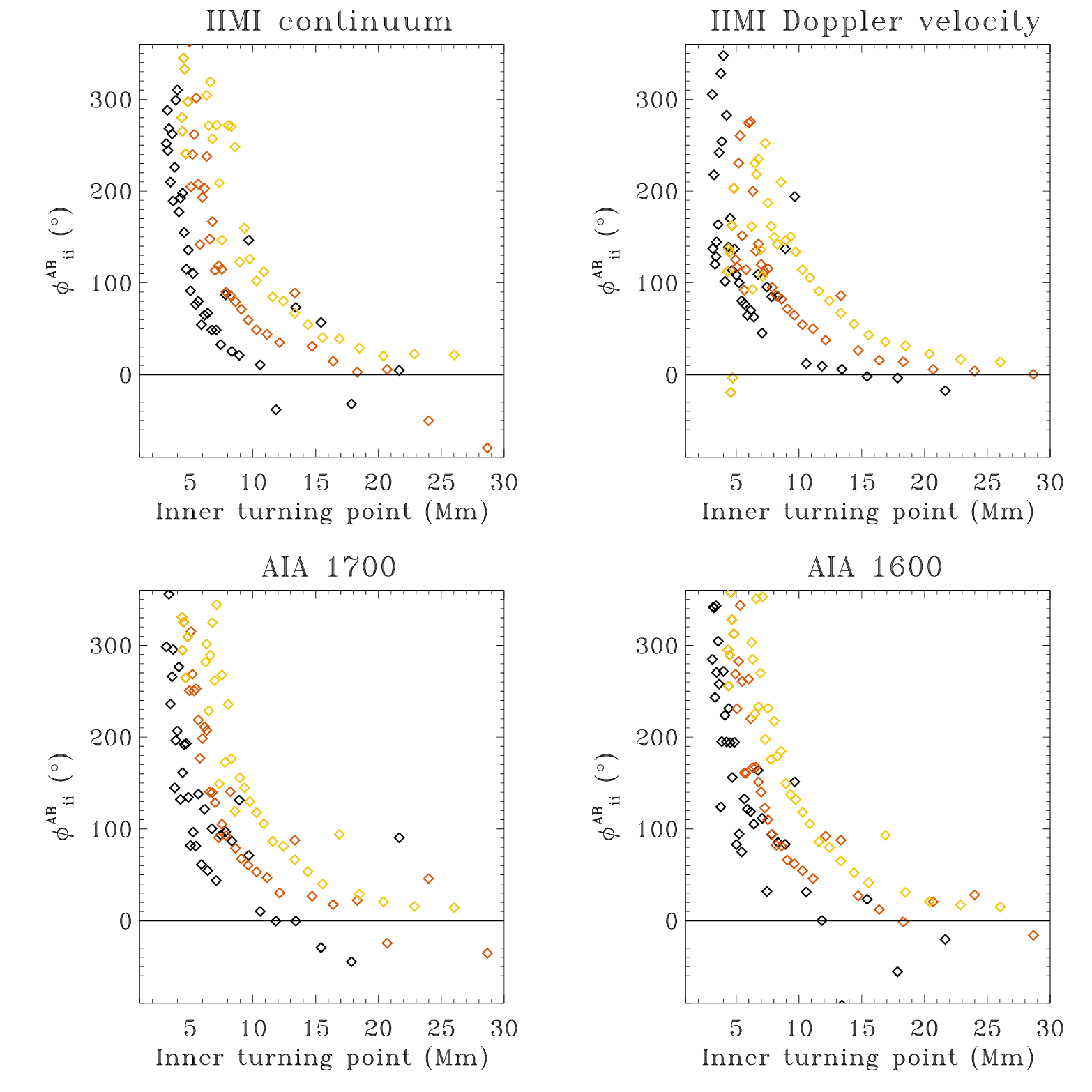}
\caption{Average of $\phi^{AB}_{ii}(k,n)$ over the six strongest sunspots as a function of the inner turning point $R_0$ of p-modes. Black diamonds are for the $n=1$ modes, the dark red diamonds are for $n=2$, and the yellow ones are for $n=3$.}
\label{sixthfigd}       
\end{figure}

Finally, we compared the respective depths over which power reduction and wave scattering occur. Figure \ref{sixthfigd} shows the phase shift for acoustic modes $n=1$, $2$, and $3$ (again averaged over the six strongest sunspots), and at the four measurement heights, as a function of inner turning points $R_0$ of the modes calculated in the ray-path approximation. $R_0$ is defined as the depth $r$ at which $c_s/R_0=\omega / \ell$. Solar model S of \mbox{J.} Christensen-Dalsgaard provided $c_s(r)$. The steep decrease of $\phi^{AB}_{ii}(k,n)$ with $R_0$ is conspicuous. This figure differs from Figure 15 of Couvidat (2013), where power reduction in sunspots was plotted as a function of $R_0$. Clearly, regions of power reduction and wave scattering are not identical, and phase scattering is a shallow surface phenomenon quickly fading with depth. Comparison of Figure \ref{sixthfigd} with Figure 15 of Couvidat (2013) also confirms that power reduction occurs somewhat deeper underneath the sunspot than phase scattering, and it occurs over a region vertically more extended. This was suggested by Cally, Crouch, and Braun (2003) who argued that the phase speed of p-modes is determined immediately below the surface where $c_s$ is lowest, while absorption (resulting from a loss of power through slow MAG waves) occurs deeper.
The region of phase scattering appears both less extended horizontally and shallower than the region of power reduction. 

Time-distance helioseismology inversions of subsurface wave speed ({\it e.g.}, Kosovichev, Duvall, and Scherrer, 2000; Couvidat, Birch, and Kosovichev, 2006) show a decrease in wave speed in the first $\approx 2.5$ Mm underneath a sunspot. If the scattering of acoustic waves resulted mainly from such a wave-speed perturbation we would expect a negative value of $\phi^{AB}_{ii}(k,n)$ for those modes whose inner turning point is closer to the surface than, or equal to, $\approx 2.5$ Mm. On the contrary, the positive values measured are difficult to reconcile with time-distance inversions ({\it e.g.} Cally, Crouch, and Braun, 2003). 

\section{Results: Ingoing - Ingoing And Outgoing - Outgoing Phase Differences}

Even though this paper focuses on outgoing-ingoing scattering phase shifts produced by sunspots, a unique advantage of contemporaneous observations at four different measurement heights is the ability to also measure $\phi^{AA}_{ij}(k,\omega)$ and $\phi^{BB}_{ij}(k,\omega)$ with $i \ne j$. Further information regarding the photospheric properties of waves can be gleaned. Figure \ref{sevfig} shows 2D maps of $\phi^{AA}_{ij}(k,\omega)$ for different $(i,j)$ pairs, averaged over the 14 sunspot datasets of our sample. Howe {\it et al.} (2012) mentioned a possible misregistration between HMI and AIA images. A pixel on an HMI image might correspond to a slightly different point at the solar surface than the same pixel on an AIA image, which would introduce spurious phase differences. To alleviate this problem, Howe {\it et al.} (2012) performed a spatial boxcar smoothing over three $0.075$ degree pixels. Here, as a test, a smoothing by $3\times3$ pixels was applied to the datacubes prior to computing their phase differences: Both smoothed and raw cubes produce results in qualitative agreement, and their computed phases differ only by a few degrees. In the rest of the paper we present phase differences from the smoothed cubes.

\subsection{Ingoing - Ingoing Phase Differences Between Two Different Atmospheric Heights}

\begin{figure}
\centering
\includegraphics[width=\textwidth]{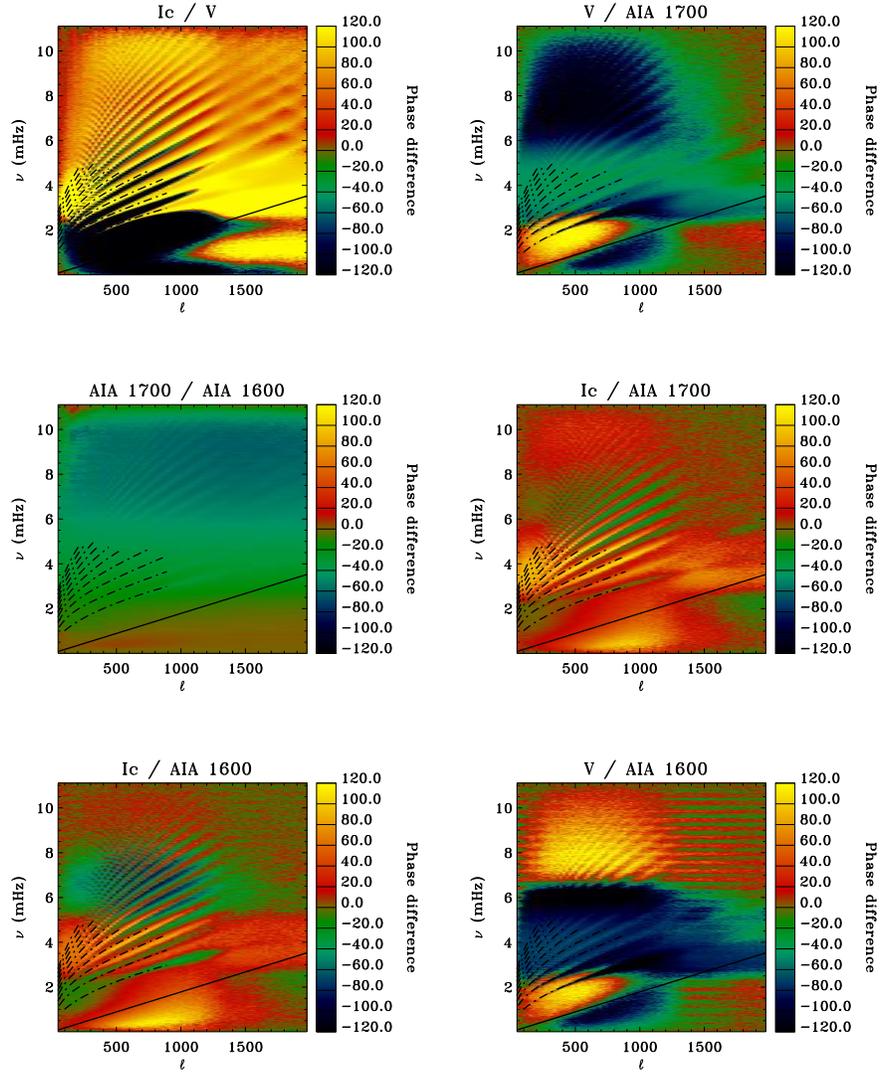}
\caption{Ingoing - ingoing phase difference $\phi^{AA}_{ij}(k,\omega)$ between two heights $i$ and $j$ in the photosphere, as a function of angular degree $\ell$ and frequency $\nu$. The black solid lines show the Lamb frequency. The curves have been smoothed and the color scale has been truncated. The dash-dotted lines show the position of some mode ridges, from a set of solar oscillation frequencies obtained with MDI Dopplergrams.}
\label{sevfig}       
\end{figure}
No correction to the phase shifts is needed with  $\phi^{AA}_{ij}(k,\omega)$, resulting in less noisy values than the shifts measured in previous sections. 

The upper left panel of Figure \ref{sevfig} shows the Ic/V phase difference, {\it i.e.} the phase of V minus the phase of Ic with our definition of Section 2, where Ic is the continuum intensity and V the Doppler velocity. This Ic/V phase diagram is very similar to Figure (2c) of Deubner {\it et al.} (1990) (the sign is reversed because with HMI Dopplergrams the velocity is positive for a downward motion, {\it i.e.} a redshift from the observer).
There is a clear discontinuity in $\phi^{AA}_{ij}(k,\omega)$ at $\nu \approx 2-2.5$ mHz, and a two-regime pattern: A continuum background with $\phi^{AA}_{ij}(k,\omega) < 0$ for frequencies below those of acoustic modes and that forms a plateau regime extending in-between the f- and p-mode ridges; and the mode ridges with $\phi^{AA}_{ij}(k,\omega) > 0$. 
At low $\ell$ (here $\ell = 23.8$), $\phi^{AA}_{ij}(k,\omega)$ decreases from $\phi^{AA}_{ij}(k,\omega) \approx 0^{\circ}$ at $\nu \approx 0.2$ mHz to $\phi^{AA}_{ij}(k,\omega) \approx -50^{\circ}$ at $\nu=2$ mHz. It then increases to reach $\phi^{AA}_{ij}(k,\omega) \approx 100^{\circ}$ at $\nu \approx 3.3$ mHz, with the zero-crossing occuring at $\nu \approx 2.2$ mHz. Finally, $\phi^{AA}_{ij}(k,\omega)$ decreases for $\nu > 3.3$ mHz, and stabilizes at $\phi^{AA}_{ij}(k,\omega) \approx 20^{\circ}$. 
The peak values of $\phi^{AA}_{ij}(k,\omega)$ increase (in absolute value) with $\ell$, to reach $\phi^{AA}_{ij}(k,\omega) \approx -155^{\circ}$ for the background regime below $\nu=2$ mHz, and $\phi^{AA}_{ij}(k,\omega) \approx 160^{\circ}$ for the p-mode ridges. The zero crossing of $\phi^{AA}_{ij}(k,\omega)$ also appears to increase with $\ell$. There is a smooth transition from p-mode ridges to pseudo-mode ridges. 
In an adiabatic atmosphere the phase difference in p-mode ridges is expected to be $90^{\circ}$: The higher value observed here probably results from non-adiabatic effects, especially radiative damping which is very strong in the lower layers of the photosphere ({\it e.g.} Mihalas and Toomre, 1982). However, the peak $\phi^{AA}_{ij}(k,\omega)$ values do not coincide with the peak power in the mode ridges (power computed from Doppler velocity).

The position of the Lamb frequency $\omega_L$ defined as $\omega_L = c_s k$, where $c_s=7.8$ km s$^{-1}$ is the average sound speed in the photosphere, is shown on each panel of Figure \ref{sevfig}. A clear shift in phase difference is observed on each side of $\omega_L$.
Deubner {\it et al.} (1990) suggested that the $\phi^{AA}_{ij}(k,\omega) < 0$ background results from a continuum of downward scattered waves, an idea supported by the theoretical work of Marmolino {\it et al.} (1993): The temperature stratification of the photosphere creates this continuum of downward reflected waves.
For $\nu > 5.3$ mHz (acoustic cut-off frequency), in the propagating-wave regime, the phase difference decreases with $\nu$, as expected from upward propagating waves (Marmolino and Severino, 1991). 

\begin{figure}
\centering
\includegraphics[width=\textwidth]{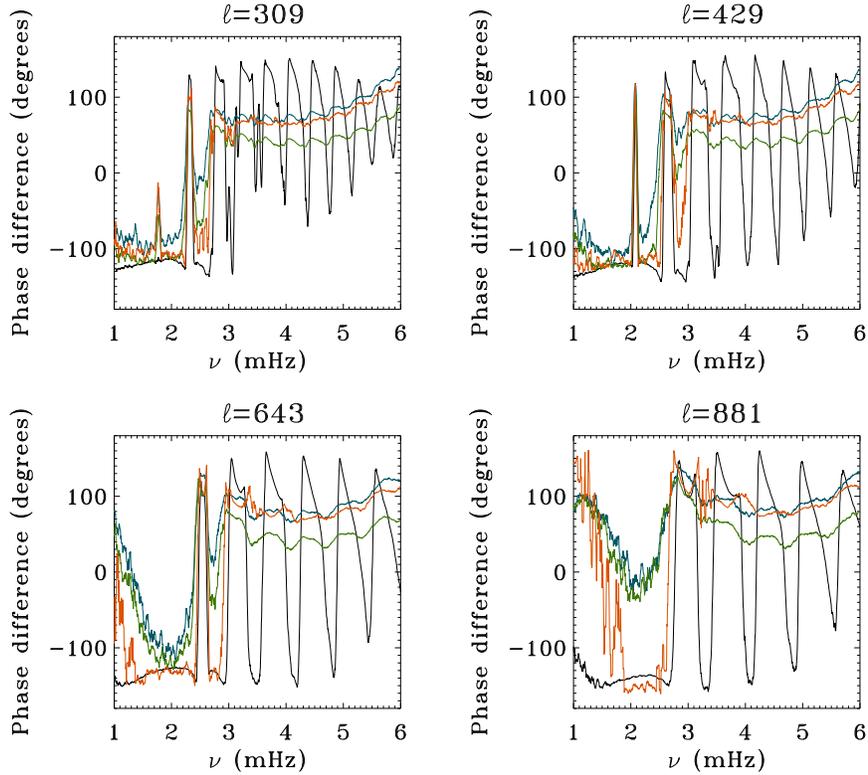}
\caption{Cuts in the ingoing - ingoing phase difference $\phi^{AA}_{ij}(k,\omega)$ between two heights $i$ and $j$ in the photosphere, as a function of frequency $\nu$ and for four angular degrees $\ell$. The black line is for the Ic/V phase difference, the red line is for the core/V phase difference, the green line is for the AIA 1700/V phase difference, and the blue line is for the AIA 1600/V phase difference. The curves have been smoothed.}
\label{figjb}       
\end{figure}

The asymmetry of p-mode peaks is the opposite of the one visible on Figure 1 of Oliviero {\it et al.} (1999) and based on {\it Global Oscillation Network Group} (GONG) data: The authors observed steeper slopes on the high-frequency side of the mode peaks, while we observe a steeper slope on the low-frequency side (Figure \ref{figjb}). This difference is due to the fact that Oliviero {\it et al.} (1999) define the velocity as positive for an upward motion.

On Figure \ref{figjb} the cuts of the Ic/V phase difference are compared with core/V, AIA 1700/V (the opposite of upper right panel of Figure \ref{sevfig}), and AIA 1600/V (the opposite of the lower right panel). The line core is the difference between HMI continuum intensity and HMI linedepth: It is an intensity image formed higher in the photosphere than the Doppler velocity. Note that the core/V phase difference was computed for only one of our datasets and is not an average, hence its noisier aspect. Two main features are noticeable on the figure: The background regime (plateau regime of Deubner {\it et al.}, 1992) below $\nu \approx 2.5$ mHz characterized by $\phi^{AA}_{ij}(k,\omega)<0$ for Ic/V and most of core/V, turns positive as $\ell$ increases for AIA 1700/Ic and AIA 1600/Ic; and the negative background regime in the inter-ridges of the Ic/V phase differences becomes positive for core/V, AIA 1700/V, and AIA 1600/V. Moreover the inter-ridge phase difference is larger for AIA 1600/Ic than for AIA 1700/Ic. This corroborates some of the findings of Deubner {\it et al.} (1992), where these authors noticed that the plateau regime below the p-mode ridges and for $\ell>700$ turns from negative in the lower photosphere to positive above the temperature minimum, in the chromosphere, while it stays negative at lower $\ell$. This is an example of how AIA signals show some sensitivity to the lower chromosphere. In the p-mode ridges, the phase difference is higher than 90$^{\circ}$ for Ic/V but lower for the other phase diagrams: These variations are likely partly due to change in radiative damping with height.

On the upper right panel of Figure \ref{sevfig} (the V/AIA 1700 phase difference), and on the lower right panel (the V/AIA 1600 phase difference), the discontinuity around $\nu=2$ mHz occurs at a somewhat lower frequency than for the Ic/V phase difference (closer to $\nu=2$ mHz than $2.2$ mHz).
The V/AIA 1600 phase difference also shows a second sharp discontinuity starting at $\nu \approx 6.5$ mHz, which is not visible on the V/AIA 1700 diagram. This discontinuity occurs in the domain of propagating waves and was observed by Deubner, Waldschik, and Steffens (1996), who explained it with a multicomponent wavefield: Upward propagating acoustic waves and two resonance modes trapped in the chromosphere. However, these authors mentioned an $180^{\circ}$ discontinuity at 8 mHz, while here this discontinuity occurs mostly between $6.5$ and 7 mHz.
The presence of such a phase jump supports the idea that the AIA 1600 \AA\ sensitivity function extends higher into the chromosphere than the three other datasets. Overall, results in this section points to a significant sensitivity of the AIA datasets to the lower chromosphere.

$\phi^{AA}_{ij}(k,\omega)$ computed for Ic/AIA 1700 has positive values in the f- and p-mode ranges, and in the background below the Lamb frequency $\omega_L$. The background value decreases with $\nu$ to zero for $\nu$ smaller than the f-mode frequencies, and increases again at $\nu \ge 6.5$ mHz.
The phase difference for Ic/AIA 1600 is similar, except that the background reaches higher values below $\omega_L$ and exhibits a steeper decrease with $\nu$. This is consistent with the phase difference of AIA 1700/AIA 1600 being negative for the background and p-mode ridges. Indeed, $\phi^{AA}_{ij}(k,\omega)$ for AIA 1700/AIA 1600 shows a linear decrease with $\nu$ (for roughly $\nu \le 6.5$ mHz) at all $\ell$: The phase difference goes from $\phi^{AA}_{ij}(k,\omega) = 0$ at $\nu=0$ to $\phi^{AA}_{ij}(k,\omega) \approx -50^{\circ}$ at $\nu \approx 6.5$ mHz, and this decrease seems to be a weak function of $\ell$. Krijger {\it et al.} (2001) interpreted this behaviour as resulting from radiative damping (in the 5-min oscillation range), and from upward acoustic-wave propagation at higher frequencies. The presence of radiative damping gives oscillation modes a mixed character: they are not purely standing waves anymore but are also partly propagating. Note that  Krijger {\it et al.} (2001) plotted the phase of TRACE 1700 minus the phase of TRACE 1600 on their Figure 18, while we plotted the opposite.

\subsection{Ingoing - Ingoing minus Outgoing - Outgoing Phase Differences At Two Different Atmospheric Heights}

\begin{figure}
\centering
\includegraphics[width=\textwidth]{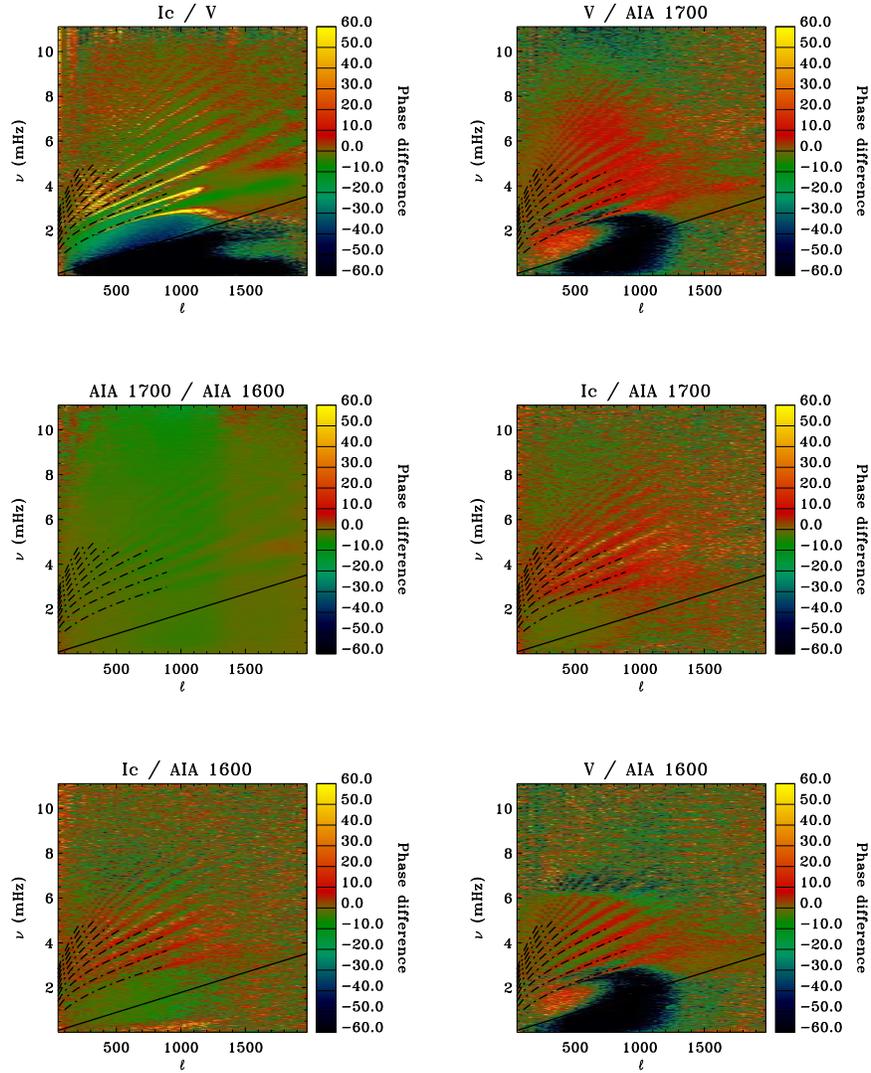}
\caption{Ingoing - ingoing phase difference $\phi^{AA}_{ij}(k,\omega)$ minus outgoing - outgoing  phase difference $\phi^{BB}_{ij}(k,\omega)$ between two heights $i$ and $j$ in the photosphere, as a function of angular degree $\ell$ and frequency $\nu$. The black solid lines show the Lamb frequency. The curves have been smoothed and the color scale has been truncated. The dash-dotted lines show the position of some mode ridges, from a set of solar oscillation frequencies obtained with MDI Dopplergrams.}
\label{figj}       
\end{figure}

Finally, we conclude this work by studying the sunspot impact on wave phases in a different manner from Section 4. The $\phi^{AA}_{ij}(k,\omega)-\phi^{BB}_{ij}(k,\omega)$ phase differences are plotted on Figure \ref{figj}. These differences arise most likely from the outgoing waves having crossed the sunspots and they highlight the magnetic-field impact on phase diagrams. On the Ic/V phase diagram (upper left panel), the waves below the Lamb modes are strongly affected by the field: This is likely due to the partial inhibition of convection in sunspots. Conversely the plateau regime above $\nu=2.5$ mHz is not perturbed. The fact that the oscillation-mode ridges are visible on the Ic/V diagram might be partly due to a small change in ridge positions between ingoing and outgoing waves (see Couvidat, 2013): Indeed $\phi^{AA}_{ij}(k,\omega)-\phi^{BB}_{ij}(k,\omega)$ inside the ridges is close to zero, but it is positive at the low-frequency edge of the ridges and negative at their high-frequency edge. It could also be partly due to a difference in asymmetry (the ``shark-fin'' shape) between the peaks of $\phi^{AA}_{ij}(k,\omega)$ and $\phi^{BB}_{ij}(k,\omega)$.
 
The V/AIA 1600 phase diagram shows that the plateau regime above $\nu=2.5$ mHz is unchanged between ingoing and outgoing waves, while $\phi^{AA}_{ij}(k,\omega)$ is $\approx 10^{\circ}$ larger than $\phi^{BB}_{ij}(k,\omega)$ in the p-mode ridges. The AIA 1700/AIA 1600 phase diagram shows a systematic offset relatively independent of $\ell$ and $\omega$ of $\approx -3^{\circ}$ between ingoing and outgoing modes (with $\phi^{BB}_{ij}(k,\omega)$ larger than $\phi^{AA}_{ij}(k,\omega)$), showing that the magnetic field affects both propagating and evanescent waves (for instance by lowering the acoustic cut-off frequency and acting on radiative damping).
The V/AIA 1700 and V/AIA 1600 phase diagrams, like the Ic/V one, show a marked impact of magnetic fields in the convective domain (supergranulation and granulation). On the V/AIA 1600 phase difference, the discontinuity at $\nu \approx 6.5$ mHz observed in the previous section is again visible.
On the Ic/AIA 1600 diagram, the plateau regime has a phase $\phi^{AA}_{ij}(k,\omega)$ smaller by $5-10^{\circ}$ than $\phi^{BB}_{ij}(k,\omega)$. 

More interesting, the power spectral densities of the Ic/AIA 1600 cross-spectra $AA_{ij}(k,\omega)$ and $BB_{ij}(k,\omega)$ clearly show the influence of acoustic glories (haloes): The power of $AA_{ij}(k,\omega)$ is smaller in the frequency range $\nu=4.5-5.5$ mHz (especially for $\ell$ in the range 100-300) than the power of $BB_{ij}(k,\omega)$. Therefore, there is power enhancement. At lower frequencies, the power reduction produced by sunspots is also visible. However, on the $\phi^{AA}_{ij}(k,\omega)-\phi^{BB}_{ij}(k,\omega)$ phase diagram of the Ic/AIA 1600 cross-spectrum (lower left panel of Figure \ref{figj}) no discontinuity or significant change is noticeable in the frequency range where acoustic haloes are most obvious: This supports the observation in Section 4 that haloes do not seem to impact significantly the phases of outgoing waves.

\section{Conclusion}

This study continued the work started in Couvidat (2013) and that takes advantage of high quality data provided at different measurement heights by the HMI and AIA instruments onboard the SDO satellite. Scattering phase shifts of acoustic waves crossing sunspots, and phase differences of waves between two measurement heights have been computed using an Hankel-Fourier transform. Following Braun (1995) and due to the high noise level on individual measurements, all the phase shifts presented here have been averaged to some degree, based on a vector-average scheme rather than a simple algebraic average: both schemes return quantitatively different values.
Four types of datasets were utilized: HMI continuum intensity, HMI Doppler velocity, AIA 1700 \AA\ intensity, and AIA 1600 \AA\ intensity. This makes it possible to span the entire photosphere and reach the lower chromosphere.
Outgoing-ingoing phase shifts show surprisingly little dependence on atmospheric height (after quiet-Sun results are subtracted). This indicates that scattering by sunspots occurs primarily very low at, or below, the continuum level with little influence of the magnetic field above. This contrasts with the acoustic power reduction of waves crossing sunspots, which was significantly dependent on measurement height. Power reduction itself occurs deeper underneath the surface than wave scattering, but other phenomena, like acoustic glories (haloes) impact acoustic power reduction coefficients above the surface while they appear to have little influence on phase shifts. The ingoing-ingoing minus outgoing-outgoing phase differences for the Ic/AIA 1600 cross-spectra support this finding. When considering only the strongest sunspots in our sample, we obtained scattering phase shifts very similar to those measured by Braun (1995) and we managed to measure them at slightly larger $\ell$.
These outgoing-ingoing phase shifts show a strong dependence on the LOS magnetic flux of sunspots, as well as on their surface areas. When plotted as a function of sunspot intensity contrasts or surface sound speeds, the shifts show little variation at first with the decrease of these quantities, but then exhibit a steep rise as some threshold is reached.
The drop in phase shift with impact parameter $m/k$ favors the idea that wave scattering occurs in an horizontal region smaller than the region of power reduction but slightly larger than the sunspot. This region also appears to be centered closer to the surface than the power-reduction region and to have a smaller vertical extension. Indeed, the steep decrease in phase shift with inner turning point of acoustic waves contrasts strongly with the curves obtained for power reduction in Couvidat (2013). The model of Cally, Crouch, and Braun (1993) of mode conversion in sunspots convincingly explains this result.
The figures of scattering phase shifts as a function of LOS magnetic flux, of sunspot radius, of lower turning point $R_0$, and of intensity contrast, should prove useful to refine and adjust the parameters of theoretical models of wave scattering by sunspots. Several models have been developed in the past, but they were all tested against the only scattering phase shift dataset available so far with the Hankel transform (Braun, 1995).

Ingoing-ingoing (or outgoing-outgoing) phase differences between two different photospheric heights were also measured and exhibit a complex behavior. The Ic/V phase difference is in agreement with earlier studies based on other instruments. The AIA 1600 \AA\ intensities (and, to a lesser extent, the AIA 1700 \AA\ ones) show a clear influence from chromospheric signals: This influence is so significant that it cannot be ignored when using AIA 1600 \AA\ for helioseismic studies. Finally, It was also shown that the magnetic field of sunspots has an impact in the supergranulation-granulation domain, and alters the phase difference diagrams in the entire ($k,\omega$) range studied. 

\begin{acks}
This work was supported by NASA Grant NAS5-02139 (HMI). The data used here are courtesy of NASA/SDO and the HMI and AIA science teams. The author thanks M.C. Rabello-Soares for providing him with a set of p-mode oscillation frequencies from the MDI instrument. The author also thanks the anonymous referee for useful comments and suggestions.
\end{acks}

\end{article}

\begin{thebibliography}{3}
\bibitem{} Borrero, J.M., Ichimoto, K.: 2011, {\it Living Reviews in Solar Physics} {\bf 8}, 4.
\bibitem{} Braun, D.C., Duvall, T.L., Jr., Labonte, B.J.: 1987, {\it Astrophys. J.} {\bf 319}, 27.
\bibitem{} Braun, D.C., Duvall, T.L., Jr., Labonte, B.J.: 1988, {\it Astrophys. J.} {\bf 335}, 1015.
\bibitem{} Braun, D.C., Duvall, T.L., Jr., Labonte, B.J., Jefferies, S.M., Harvey, J.W., Pomerantz, M.A.: 1992, {\it Astrophys. J.} {\bf 391}, 113.
\bibitem{} Braun, D.C.: 1995, {\it Astrophys. J.} {\bf 451}, 859.
\bibitem{} Cally, P.S., Crouch, A.D., Braun, D.C.: 2003, {\it Mon. Not. R. Astron. Soc.} {\bf 346}, 381.
\bibitem{} Chou, D.-Y., Chen, C.-K.: 1993, in Global Oscillation Network Group Report Number 10: 1993 Annual GONG Meeting Abstracts (Tucson), 26
\bibitem{} Couvidat, S., Birch, A.C., Kosovichev, A.G.: 2006,  {\it Astrophys. J.} {\bf 640}, 516.
\bibitem{} Couvidat, S.: 2013, {\it Solar Phys.} {\bf 282}, 15.
\bibitem{} Crouch, A.D., Cally, P.S., Charbonneau, P., Braun, D.C., Desjardins, M.: 2005, {\it Mon. Not. R. Astron. Soc.} {\bf 363}, 1188. 
\bibitem{} Deubner, F.-L., Fleck, B., Marmolino, C., Severino, G.: 1990, {\it Astron. Astrophys.} {\bf 236}, 509.
\bibitem{} Deubner, F.-L., Fleck, B., Schmitz, F., Straus, T.: 1992, {\it Astron. Astrophys.} {\bf 266}, 560.
\bibitem{} Deubner, F.-L., Waldschik, T., Steffens, S.: 1996, {\it Astron. Astrophys.} {\bf 307}, 936.
\bibitem{} Fan, Y., Braun, D.C., Chou, D.-Y.: 1995, {\it Astrophys. J.} {\bf 451}, 877.
\bibitem{} Gordovskyy, M., Jain, R., Thompson, M.J.: 2006, Proceedings of SOHO 18/GONG 2006/HELAS I, Beyond the spherical Sun (ESA SP-624). 7-11 August 2006, Sheffield, UK. Editor: Karen Fletcher. Scientific Editor: Michael Thompson, Published on CDROM, p.14.1
\bibitem{} Gordovskyy, M., Jain, R.: 2007a, {\it Astron. Nach.} {\bf 328}, 309.
\bibitem{} Gordovskyy, M., Jain, R.: 2007b, {\it Astrophys. J.} {\bf 661}, 586.
\bibitem{} Hill, F., Jain, K., Tripathy, S., Kholikov, S., Gonzalez Hernandez, I., Leibacher, J., {\it et al.}: 2011, American Astronomical Society, SPD meeting 42, 21.11; Bulletin of the American Astronomical Society, Vol. 43.
\bibitem{} Howe, R., Hill, F., Komm, R., Broomhall, A.-M., Chaplin, W.J., Elsworth, Y.: 2011, {\it Journal of Physics: Conference Series} {\bf 271}, 012058.
\bibitem{} Howe, R., Jain, K., Bogart, R.S., Haber, D.A., Baldner, C.S.: 2012, {\it Solar Phys.}, submitted
\bibitem{} Kosovichev, A.G., Duvall, T.L., Jr., Scherrer, P.H.: 2000, {\it Solar Phys.} {\bf 192}, 159.
\bibitem{} Krijger, J.M., Rutten, R.J., Lites, B.W., Straus, T., Shine, R.A., Tarbell, T.D.: 2001, {\it Astron. Astrophys.} {\bf 379}, 1052.
\bibitem{} Lemen, J.R., Title, A.M., Akin, D.J., Boerner, P.F., Chou, C., Drake, J.F., {\it et al.}: 2012, {\it Solar Phys.} {\bf 275}, 17.
\bibitem{} Marmolino, C., Severino, G.: 1991, {\it Astron. Astrophys.} {\bf 242}, 271.
\bibitem{} Marmolino, C., Severino, G., Deubner, F.-L., Fleck, B.: 1993, {\it Astron. Astrophys.} {\bf 278}, 617.
\bibitem{} Mihalas, B. W., Toomre, J.: 1982, {\it Astrophys. J.} {\bf 263}, 386.
\bibitem{} Oliviero, M., Severino, G. Straus, T., Jefferies, S.M., Appourchaux, T.: 1999, {\it Astron. Astrophys.} {\bf 516}, L45.
\bibitem{} Rabello-Soares, M.C., Korzennik, S.G., Schou, J.: 2008, {\it Solar Phys.} {\bf 257}, 197.
\bibitem{} Rajaguru, S.P., Couvidat, S., Sun, X., Hayashi, K., Schunker, H.: 2012, {\it Solar Phys.}, {\it accepted}, DOI: 10.1007/s11207-012-0180-9
\bibitem{} Scherrer, P.H., Bogart, R.S., Bush, R.I., Hoeksema, J.T., Kosovichev, A.G., Schou, J., {\it et al.}: 1995, {\it Solar Phys.} {\bf 162}, 129.
\bibitem{} Schou, J., Scherrer, P.H., Bush, R.I., Wachter, R., Couvidat, S., Rabello-Soares, M.C., {\it et al.}: 2012, {\it Solar Phys.} {\bf 275}, 229.
\end{thebibliography}
\end{document}